\documentstyle[12pt]{article}
\input mssymb
\begin{document}
\begin{center}
{\Large\bf Kinks and Time Machines}\\
\end{center}
\vspace*{0.3cm}

\begin{center}
{\large Andrew Chamblin, G.W. Gibbons, Alan R. Steif}\\
{\it Department of Applied Mathematics and Theoretical Physics,}\\
{\it Silver Street, Cambridge CB3 9EW, England.}\\
\end{center}
\vspace*{0.6cm}

\vspace*{0.6cm}{\small We show that it is not possible to smooth out
the metric on the Deutsch-Politzer time machine to obtain an
everywhere non-singular asymptotically flat Lorentzian metric.}\\
\vspace*{0.6cm}

{There has been interest recently in asymptotically flat spacetimes
containing closed timelike curves (CTCs). One widely discussed example
of such a spacetime (sometimes referred to as the Deutsch-Politzer
spacetime) is constructed as follows [1, 2]. One takes ordinary
$n$-dimensional Minkowski spacetime with inertial coordinates
$(t,{\mbox{\bf x}})$ and deletes two spatial balls, i.e., one removes
the points}
\[
{t = 0, ~|{\mbox{\bf x}}| < 1}
\]
{and}
\[
{t = 1, ~|{\mbox{\bf x}}| < 1.}
\]
{\noindent One now identifies the upper side of the earlier ball with
the lower side of the later ball and the upper side of the later ball
with the lower side of the earlier point. That is, one identifies
points $t = {\epsilon}, ~|{\mbox{\bf x}}| ~{\leq}~ 1$ with points $t =
1 - {\epsilon}, ~|{\mbox{\bf x}}| ~{\leq}~ 1$ and points $t =
-{\epsilon}, ~|{\mbox{\bf x}}| ~{\leq}~ 1$ with points $t = 1 ~+~
{\epsilon}, ~|{\mbox{\bf x}}| ~{\leq}~ 1$ having the same ${\mbox{\bf
x}}$ values, where ${\epsilon}$ is an infinitesimal positive quantity.
The metric is, as we shall formally demonstrate later, singular at the
branch points $t = 0, ~t = 1, ~|{\mbox{\bf x}}| = 1$. It is clear that
as a manifold, we can smooth out the singularities at $t = 0,
{}~|{\mbox{\bf x}}| = 1$, to obtain a smooth manifold with the topology
of ${\Bbb R}^{n}$ with a handle attached, i.e., which is homeomorphic
to $S^{1} ~{\times}~ S^{n - 1} - {\{}{\mbox{pt}}{\}}$, where the point
corresponds to infinity. It is less clear whether one can smooth out
the spacetime metric to obtain a smooth time-orientable Lorentzian
spacetime with an everywhere non-singular Lorentz metric, $g$, which
is asymptotically flat. In fact one cannot: no matter how one tries,
the metric must always contain singularities. The point of this note
is to give a simple proof of this fact in the case that $n$ is even.
This covers the two most interesting cases; $n = 4$, the physical
case, and $n = 2$, which provides a simple toy model.

Our proof makes use of the properties of a {\it gravitational kink}
which we have developed previously for dealing with topological
problems involving Lorentzian metrics [3, 4, 5]. We begin by
surrounding the region of CTCs, $0 < t < 1, ~|{\mbox{\bf x}}| < 1$,
with a large $(n - 1)$-sphere, $S_{\infty}$,}
\[
{t^{2} ~+~ |{\mbox{\bf x}}|^{2} = R^{2}}
\]
{where $R$ is much larger than the size of the time machine. We now
have a compact manifold, $M$, with boundary ${\partial}M ~{\cong}~
S^{n - 1}$.

The necessary and sufficient condition that any compact manifold, $M$
with boundary ${\partial}M$ admit a Lorentz metric $g_{L}$ is that}
\begin{equation}
{{\chi}[M] = {\mbox{kink}}({\partial}M; g_{L})}
\end{equation}
{where ${\chi}[M]$ is the Euler characteristic of $M$ and
${\mbox{kink}} ({\partial}M; g_{L})$ is the kink number of the Lorentz
metric $g_{L}$ with respect to the boundary ${\partial}M$ [3]. The
definition of the kink number ${\mbox{kink}} ({\partial}M; g_{L})$ is
as follows. One introduces on $M$ an auxillary Riemannian metric
$g_{R}$ and then diagonalises $g_{L}$ with respect to $g_{R}$. Because
$g_{L}$ is assumed to be time-orientable, the field of eigenvectors
with negative eigenvalue provides an everywhere non-vanishing vector
field ${\mbox{\bf V}}$, which may be assumed to be normalised to have
unit length with respect to the Riemannian metric $g_{R}$. The kink
number ${\mbox{kink}}({\partial}M; g_{L})$ is now the algebraic number
of times that ${\mbox{\bf V}}$ coincides with the inward pointing unit
normal ${\mbox{\bf n}}$ of ${\partial}M$. More precisely, ${\mbox{\bf
V}}$ and ${\mbox{\bf n}}$ provide two global sections of the bundle
$S({\partial}M)$ of unit 4-vectors over ${\partial}M$. Then two
sections intersect generically in an isolated number of points which
may be assigned a sign depending upon the orientations of the
sections. One then counts the number of points with regard to sign.
The orientation conventions are such that if one reverses the sign of
the normal, then the sign of the kink number changes. Reversing the
sign of the vector field ${\mbox{\bf V}}$, however, leaves the kink
number unchanged.

In two spacetime dimensions, one may characterise the kink number in
terms of the more familiar winding number. Let ${\theta}$ be the angle
through which ${\mbox{\bf V}}$ must be rotated counter-clockwise to
coincide with ${\mbox{\bf n}}$. The kink number is the total change in
${\frac{\theta}{2{\pi}}}$ as the boundary ${\partial}M$ is traversed
with the inward normal ${\mbox{\bf n}}$ pointing left. Changing the
sign of ${\mbox{\bf V}}$ will merely change ${\theta}$ to ${\theta}
{}~+~ {\pi}$ everywhere on the boundary and thus leave invariant the
total change in ${\theta}$ and hence the kink number. However,
changing the sign of ${\mbox{\bf n}}$ will reverse the kink number
because now the boundary must be traversed in the opposite direction
(see Figure 1).

Applying equation (1) to ordinary Minkowski spacetime inside the large
sphere $S_{\infty}$ (or calculating directly) shows that}
\begin{equation}
{{\mbox{kink}}(S_{\infty}; g_{L}) = +1 .}
\end{equation}
{However, one has}
\begin{equation}
{{\chi}(S^{1} ~{\times}~ S^{n - 1} - {\{}{\mbox{pt}}{\}}) = (-1)^{n -
1} .}
\end{equation}
{Thus equation (1) cannot be satisfied for even $n$. More generally,
we could consider a spacetime with $g$ time machines. This would have
the topology of the connected sum, denoted by ${\#}$, of ${\Bbb
R}^{n}$ with $g$ handles. One has}
\begin{equation}
{{\chi}({\Bbb R}^{n} {\#} S^{1} ~{\times}~ S^{n - 1} {\#} S^{1}
{}~{\times}~ S^{n - 1} ... {\#} S^{1} ~{\times}~ S^{n - 1}, S_{\infty})
= 1 - g(1 ~+~ (-1)^{n})}
\end{equation}
{which again contradicts equation (1). We note in passing (see [5])
that these results also go through in the non-time-orientable case.

We now show directly that the branch points $|{\mbox{\bf x}}| = 1$ are
singular in the two-dimensional case. Consider a curve enclosing one
of the branch points, say $x = 1$. As in the previous discussion, we
calculate the winding number of the timelike vector field
${\frac{\partial}{{\partial}t}}$ with respect to the inward normal to
the curve.  Since the curve winds first around the point $t = 1$,
$|{\mbox{\bf x}}| = 1$ and then around $t = 0$, $|{\mbox{\bf x}}| =
1$, one sees that the kink number is two.  However, the kink number
which is obtained upon encircling a non-singular point in Minkowski
space is unity. Therefore, the spacetime is not locally Minkowskian at
the branch point. This argument is readily generalised to higher
dimensions.

It is perhaps worth emphasizing that the manifold $S^{1} ~{\times}~
S^{n - 1} - {\{}{\mbox{pt}}{\}}$ does in fact admit a non-singular
Lorentz metric. For example, take the vector field which winds around
the $S^{1}$ factor. This Lorentz metric however has kink number $= -1$
and so cannot be asymptotically flat. This difference may be traced
back to the fact that the kink number changes sign on reversal of the
normal (see Figure 3).

We have shown (at least for even $n$) that one cannot construct an
asymptotically flat Deutsch-Politzer type time machine with an
everywhere time-orientable non-singular Lorentz metric. For $n = 2$,
it does not seem that one can construct a non-trivial non-singular
model at all since all one is allowed is to add on more handles, but
by equation (4), this makes things worse. In the case of
four-dimensional spacetimes, it is clear that one can construct
asymptotically flat time machines, but these must be considerably more
complicated than the Deutsch-Politzer model.

{}From the physical point of view, it is not clear to us how seriously
one should worry about the singularities in the Deutsch-Politzer
spacetime. For the purposes of discussing conceptual problems
involving CTCs, the view has been expressed to us that one may always
impose boundary conditions at these singularities that prevent
information from them affecting such issues as the unitarity of
quantum field theory on these backgrounds. What is clear to us is that
the gravitational kink concept is useful in detecting the presence of
such singularities. In this respect, the situation resembles attempts
to model topology change using a trouser-leg spacetime -- i.e., one
diffeomorphic to $S^{n}$ with balls removed and with no spacelike
boundary components.  Equation (1) shows that no non-singular Lorentz
metric is possible on the trouser-leg spacetime. If $n = 4$, one can
find a spacetime with three spacelike boundary components -- take
${\Bbb C}{\Bbb P}^{2}$ and remove three four-balls for example.
However, the resulting spacetime does not admit a spinor structure.
The results of [6] show that if $n = 4$, no such Lorentz cobordism
admits a spinor structure. On the other hand, there will exist such
cobordisms which will admit a globally defined Cliffordian {\it pin}
structure (see [5]). Put another way, by allowing our cobordisms to be
non-orientable (see [5]), we can have arbitrary {\it spacelike}
topology change while preserving the global existence of a
non-vanishing Lorentz metric and a `pinor' bundle (which we can use to
quantise the Dirac equation).

Of course, one could take the view that one must go beyond globally
defined Lorentzian metrics and consider singular metrics if one wishes
to adhere rigidly to the idea that a single spacetime metric is
involved in topology change. Perhaps the least drastic modification of
the globally defined Lorentzian metric concept is that of a metric
with regions of different spacetime signature.}\\
\vspace*{0.6cm}

\begin{center}
{\bf ACKNOWLEDGMENTS}
\end{center}

{Eternal gratitude from A.C. goes to Jo Chamblin (Piglit) for loving
support and help with the preparation of this paper. A.S. acknowledges
financial support from the S.E.R.C., and A.C. is supported by N.S.F.
Graduate Fellowship No.  RCD-9255644.}\\
\vspace*{0.6cm}

{\noindent [1] D. Deutsch, Phys. Rev. {\bf D44}, 3197 (1991).}\\
\\
{\noindent [2] H.D. Politzer, Phys. Rev. {\bf D46}, 4470 (1992).}\\
\\
{\noindent [3] G.W. Gibbons and S.W. Hawking, Phys. Rev. Lett. {\bf
69}, 1719 (1992).}\\
\\
{\noindent [4] A. Chamblin and R. Penrose, Twistor Newsletter {\bf
34}, 13 (1992).}\\
\\
{\noindent [5] A. Chamblin, J. Geom. Phys. (to appear) (1993).}\\
\\
{\noindent [6] G.W. Gibbons and S.W. Hawking, Comm. Math. Phys. {\bf
148}, 345 (1992).}\\

\end{document}